\documentclass[useAMS,usenatbib,aas_macros]{mn2e}

\usepackage{hhline}
\usepackage{lscape}
\usepackage{graphicx}
\usepackage{colortbl}
\usepackage{amssymb}
\usepackage{amsmath}
\usepackage{natbib}
\usepackage{times}
\usepackage{aas_macros}
\usepackage{threeparttable}

\newcommand{\msun}{~\mathrm{M}_{\odot}}
\newcommand{\Jc}{\rm{J}_{\rm crit}}
\newcommand{\Jlw}{\rm{J}_{\rm LW}}
\newcommand{\zsun}{~\mathrm{Z}_{\odot}}
\usepackage{color}
\def\simpropto{\lower.2ex\hbox{$\; \buildrel \propto \over \sim \;$}}
\def\ltsim{\lower.5ex\hbox{$\; \buildrel < \over \sim \;$}}
\def\gtsim{\lower.5ex\hbox{$\; \buildrel > \over \sim \;$}}
\def\hii{\mbox{H\,{\sc ii}}}

\begin{document}
\title[Effects of binaries on DCBH formation]{Effects of binary stellar populations on direct collapse black hole formation}
\author[B. Agarwal, et al.]{Bhaskar Agarwal$^{1}$\thanks{E-mail:
bhaskar.agarwal@uni--heidelberg.de}, Fergus Cullen$^2$, Sadegh Khochfar$^2$, Ralf S. Klessen$^1$, 
\newauthor Simon C. O. Glover$^1$, Jarrett Johnson$^3$\\
$^1$ Universit\"{a}t Heidelberg, Zentrum f\"{u}r Astronomie, Institut f\"{u}r Theoretische Astrophysik, Albert-Ueberle-Stra{\ss}e 2, 69120 Heidelberg, Germany\\
%$^2$ Department of Astronomy, 52 Hillhouse Avenue, Steinbach Hall, Yale University, New Haven, CT 06511, USA\\
$^2$ Institute for Astronomy, University of Edinburgh, Royal Observatory, Edinburgh, EH9 3HJ\\
$^3$ X Theoretical Division, Los Alamos National Laboratory, Los Alamos, NM 87545, USA\\}

% ----------------------------------------------------------------

\date{00 Jun 2014}
\pagerange{\pageref{firstpage}--\pageref{lastpage}} \pubyear{0000}
\maketitle

\label{firstpage}

% ----------------------------------------------------------------
\begin{abstract}

The critical Lyman--Werner flux required for direct collapse blackholes (DCBH) formation, or $\Jc$, depends on the shape of the irradiating spectral energy distribution (SED). The SEDs employed thus far have been representative of {{realistic}} single stellar populations. We study the effect of binary stellar populations on the formation of DCBH, as a result of their contribution to the Lyman--Werner radiation field. Although binary populations with ages $>$ 10 Myr yield a larger LW photon output, we find that the corresponding values of $\Jc$ can be up to 100 times higher than single stellar populations. We attribute this to the shape of the binary SEDs as they produce a sub--critical rate of H$^-$ photodetaching 0.76 eV photons as compared to single stellar populations, reaffirming the role that H$^-$ plays in DCBH formation. This further corroborates the idea that DCBH formation is better understood in terms of a critical region in the H$_2$--H$^-$ photo--destruction rate parameter space, rather than a single value of LW flux.
 \end{abstract}
% ----------------------------------------------------------------


%%%%%%%%%%%%%%%%%%%%%%%%%%%%%%%%%%%%%%%%%%%%%%%%%%%%%%%%%%%%%%%%%%%%%%%%%%
%%%%%%%%%%%%%%%%%%%%%%%%%%%%%%%%%%%%%%%%%%%%%%%%%%%%%%%%%%%%%%%%%%%%%%%%%%

%%%%%%%%%%%%%%%%%%%%%%%%%%%%%%%%%%%%%%%%%%
\section{Introduction}
%%%%%%%%%%%%%%%%%%%%%%%%%%%%%%%%%%%%%%%%%%

Direct collapse black holes (DCBH) have gathered much attention recently \citep{Dijkstra2014a,Ferrara14a,Agarwal12,Agarwal14,Habouzit16a} as a plausible solution to the problem of forming  billion solar mass black holes very early in cosmic history as is required to explain the existence of very luminous quasars at redshifts $z>6$. 
Pristine gas in an atomic cooling halo exposed to a critical level of Lyman--Werner (LW) radiation can rid itself of molecular hydrogen (cooling threshold $\sim 200$ K), thereby collapsing isothermally in the presence of atomic hydrogen (cooling threshold $\sim 8000$ K). This leads to a Jeans mass threshold of $10^6 \msun$ at $n\sim 10^3 \rm \ cm^{-3}$, thereby allowing the entire gas mass in the halo \footnote{An atomic cooling halo, i.e. $\rm T_{vir}=10^4 \ K$ corresponds to a $\rm M_{DM} \approx 10^7 \msun$ at $z\approx10$. If we assume that the baryon fraction in this halo is the same as the cosmological mean value, i.e. $f_b \approx 0.16$, then the baryonic mass of such a halo will be at least $10^6 \msun$} to undergo runaway collapse eventually forming a $10^{4-5} \msun$ black hole in one go \citep{Omukai:2001p128}. The collapse must withstand fragmentation into Population III (Pop III) stars, which requires the gas to get rid of its angular momentum via bars--within--bars instabilities \citep{Begelman:2006p3700}, low--spin disks \citep[e.g.][]{Bromm:2003p22,Koushiappas:2004p871,Regan08,Lodato:2006p375} or high inflow rates in turbulent medium \citep{Volonteri:2005p793,Latif:2013p3629,Schleicher:2013p3661,Borm13}.

In order for this mechanism to work, initially there must be a LW radiation field strong enough to delay Pop III star formation in a minihalo, $\rm 2000<T_{vir}\le10^4\ K$, till it reaches the atomic cooling limit of $\rm T_{vir} \ge 10^4\ K$ \citep{Machacek:2001p150,OShea:2008p41} . At this point, the flux of LW radiation illuminating the halo from nearby external stellar source(s) must be higher than a critical value $\Jc$ (conventionally written in units of $10^{-21} \rm\ erg/s/cm^2/sr/Hz$) to facilitate isothermal collapse of the pristine gas at 8000 K into a DCBH {\cite[e.g. recent simulations by ][]{2014ApJ...795..137R,2014MNRAS.445L.109I,2015MNRAS.446.2380B}}. Many previous studies of DCBH formation have adopted highly simplified prescriptions for the spectrum of this external radiation field, approximating the spectrum of a source dominated by Pop III stars as a $\rm T = 10^5$~K black body, and of a source dominated by Population II (Pop II) stars as a $\rm T=10^4$ K  black body \citep{Omukai:2001p128,Shang:2010p33,WolcottGreen:2012p3854}. However, recent studies have emphasised the need for using more realistic spectral energy distributions (SED) for these sources as the value of $\Jc$ depends on the shape of the irradiating source's SED \citep[, A16 hereafter]{Sugimura:2014p3946, Agarwal15a, Agarwal15b}. These studies employed single stellar populations to represent the SEDs of Pop II stars, generating them using publicly available single stellar synthesis codes such as {\sc{Starburst99}} \citep{Leitherer:1999p112}, {\sc{Yggdrasil}} \citep{Zackrisson11} and \citet{Bruzual:2003p3256} model.
However, in reality it is likely that a significant number of the stars will be part of binary systems. Stellar populations with significant binary fractions have higher hydrogen ionising photon yields than single stellar populations \citep[e.g.][]{2016MNRAS.456..485S,2016MNRAS.459.3614M}, and so it is plausible that accounting for their existence will lead to significant differences in the value of J$_{\rm crit}$ that we derive.

%The effect of binary stars on DCBH formation has been overlooked thus far, and given their high H ionising photon yield as compared to single stellar populations \citep[e.g.][]{2016MNRAS.456..485S,2016MNRAS.459.3614M}, we explore their impact in this study.

%%%%%%%%%%%%%%%%%%%%%%%%%%%%%%%%%%%%%%%%%%
\section{Methodology}
%%%%%%%%%%%%%%%%%%%%%%%%%%%%%%%%%%%%%%%%%%
%%%%%%%%%%%%%%%%%%%%%%%%%%%%%%%%%%%%%%%%%%
\begin{figure}
\includegraphics[width=0.7\columnwidth,angle=90,trim={1.75cm 2cm 2.25cm 2cm},clip]{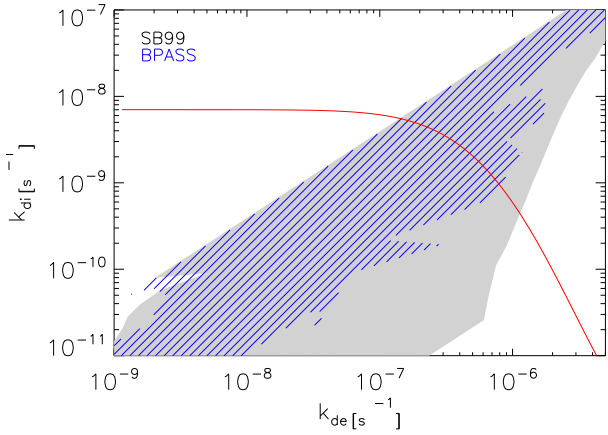}
\caption{The solid red curve is criterion for direct collapse derived described in A16, given by Eq.~\ref{eq.ratecurve}. The grey shaded region shows the range of $\rm k_{de}$ and $\rm k_{di}$ derived from SB99 stellar populations, while the blue region is the range derived from BPASS for a range of stellar populations described in Tab.~\ref{tab.stellarmodels}}
\label{fig.ratecurve}
\end{figure}
%%%%%%%%%%%%%%%%%%%%%%%%%%%%%%%%%%%%%%%%%%
We apply the framework described in A16 to SEDs generated with the stellar population synthesis code `Binary Population and Spectral Synthesis' \citep{2016MNRAS.456..485S, 2016arXiv160203790E} in its second version, BPASSv2. This is done to assess the impact of binaries on the critical LW radiation field strength required to suppress H$_{2}$ formation and enable direct collapse black hole formation.
The unique feature of the BPASSv2 models is the inclusion of massive binary star evolution which, in the context of this work, has the effect of boosting the LW photon flux at older stellar ages (see Section. 3). 

We have been motivated to consider the effects of binary star evolution by observations of local \hii \ regions which have indicated that $\gtrsim 70 \%$ of massive stars undergo a binary interaction in their lifetimes \citep[e.g.][]{Sana12}. 
Furthermore, it has been reported recently that the BPASSv2 models are better able to account for (i) the observed shape of the FUV continuum and (ii) UV + optical emission line ratios of star forming galaxies at $z \simeq 2 - 3$ \citep{2016ApJ...826..159S, 2016arXiv160802587S}, as well as the properties of massive star clusters in local galaxies \citep{2016MNRAS.457.4296W} and Pop III stars \citep{clark11sc,Greif:2012p2733,2013MNRAS.433.1094S}.
 {Given this context, it is useful to know how the presence of massive binary stars in stellar population will affect direct collapse black hole formation.
Briefly, in the BPASSv2 models, the main consequence of close binary interactions is the removal of the hydrogen envelope in primary stars, part of which accretes onto the companion secondary star resulting in its rejuvenation \citep[e.g.][]{deMink13,Podsiadlowski92}. The resulting effect on a stellar population containing a significant binary fraction is more hot-helium and Wolf-Rayet stars in the primary population, and an effective increase in the main sequence lifetimes of secondary stars. The mass transfer is also accompanied by angular momentum transfer, which causes stars to spin-up and results in a rotational mixing of layers allowing hydrogen to burn more efficiently; this effect, known as quasi-homogeneous evolution (QHE), is particularly strong at low metallicities \citep[see][]{2016arXiv160203790E,2016MNRAS.456..485S}.
The most relevant consequence of these differences on the DCBH formation scenario is that compared to single star models, the BPASSv2 binary models extend the time period over which a stellar population can emit UV photons in the LW band.}

The SED grid explored in this study is described in Tab.~\ref{tab.stellarmodels}. It is compared to the SB99 case, which we have discussed in detail in the Appendix of A16.
For the BPASSv2 models we have assumed the instantaneous burst models with ages ranging from 10$^{6-9}$ yr, a metallicity of $0.05 Z_{\odot}$  {and a 70\% binary fraction}.
In order to understand the effect of these SEDs on DCBH formation, we make the following assumptions: 
%%%%%%%%%%%%%%%%%%%%%%%%%%%%%%%%%%%%%%%%%%
\begin{table}
\caption{Summary of the stellar populations considered in this study, BPASSv2 and SB99.}
\begin{threeparttable}
\begin{tabular*}{\columnwidth}{@{\extracolsep{\fill}}l | cccc}
\hline
Instantaneous & Stellar Mass & Age $^{\rm}$ & Metallicity & IMF $^{\rm [b]}$\\ 
Burst & \tiny{($\msun$)} & \tiny{(yr)} & \tiny{($\zsun$)}  \\ 
 \\ [-1.5ex] \hline \\ [-1.5ex]
BPASSv2 &$10^{5-10}$  &$10^{6-9}$~yr &0.05 & Kroupa\\
SB99 &$10^{5-10}$  &$10^{6-9}$~yr &0.02 & Kroupa\\ [1.5ex] \hline
\end{tabular*}
\label{tab.stellarmodels}
\begin{tablenotes}
%\item[a] {Age for BPASS is given by the function $10^{(6 + 0.1n)}\ \rm yr$ where $n=[0,30]$, while for SB99 it is equispaced logarithmically between $10^6-10^9\ \rm yr$ with 1000 steps}
\item[b] {IMF of the form $\Psi(M_*) = M_*^{-\alpha} $ where $\alpha \sim 1.3$ for $0.1\leq M_* < 0.5$ and $\alpha \sim 2.35$ for $0.5\leq M_*\leq 100\ \msun$}
\end{tablenotes}
\end{threeparttable}
\end{table}
%%%%%%%%%%%%%%%%%%%%%%%%%%%%%%%%%%%%%%%%%%

\begin{enumerate}
\item The SEDs represent a galaxy of a certain age and stellar mass in a halo.
\item The DCBH formation region (in a pristine atomic cooling halo) is external to the galaxy, at an assumed separation of $5, 12, 20$ physical kpc \citep{Agarwal14}.
\item We parametrise the critical LW radaition requirement for DCBH formation in terms of the rate of photodissociation of molecular hydrogen $\rm k_{di}\ \rm (s^{-1})$, and rate of photodetachment of H$^-$, $\rm k_{de}\ \rm (s^{-1})$ where
\begin{eqnarray}
\rm k_{de} = \kappa_{de}\alpha J_{LW} \\
\rm k_{di} = \kappa_{di}\beta J_{LW}
\label{eq.destroy}
\end{eqnarray}
Here $\alpha$ and $\beta$ are rate parameters that depend on the shape of the SED (\citealt{Omukai:2001p128, Agarwal15a}; A16), {$\kappa_{\rm de} = 10^{-10} \: {\rm s^{-1}}$ and $\kappa_{\rm di} = 10^{-12} \: {\rm s^{-1}}$ are normalisation constants \citep{Agarwal15a}}, and J$_{\rm LW}$ is the mean specific intensity of the Lyman-Werner radiation field at 13.6~eV. {The latter depends on the choice of stellar population and the assumed separation between the galaxy and the atomic cooling halo.}\\

\item In A16 we showed that in our simple one-zone model of the thermal evolution of gas in the atomic cooling halo, DCBH formation occurs when the H$_{2}$ photodissociation rate exceeds a value given approximately by

\begin{equation}
\rm k_{di} \geq 10^{Aexp(\frac{-z^2}{2}) + D}\ (\rm s^{-1}),
\label{eq.ratecurve}
\end{equation}
where $z=\frac{\log_{10}(\rm k_{de}) - B}{C}$ and $A = -3.864,\ B = -4.763,\ C = 0.773$, and $D = -8.154$, for $\rm k_{de}< 10^{-5}\rm \ s^{-1}$.
\end{enumerate}
{Recently, \cite{Wolcott2017} have also advocated the usage of such a critical curve, albeit minor differences with A16 due to their computational setup.}

% {The k$_{\rm de}$--k$_{\rm di}$ parameter space (computed at 5 kpc) explored here is shown in Fig.~\ref{fig.ratecurve}: SB99 in grey and BPASSv2 in blue. The red curve depicts Eq.~\ref{eq.ratecurve}. As explained in A16, the coloured region above the red curve is where our Enzo runs return isothermal collapse of pristine gas, in an atomic cooling halo, at $\sim 8000$~K, or in other words, regions above the red curve allow for DCBH formation. Note that higher k$_{\rm de}$ values are representative of older stellar populations, whereas a higher k$_{\rm di}$ is representative of younger stellar populations. This is due to the nature of $\alpha$ and $\beta$ that govern Eqs.~\ref{eq.destroy} \citep[A16]{Agarwal15a}.}

{By computing $k_{\rm di}$ and $k_{\rm de}$ for each different SED in our two grids of models and each different separation, we can therefore determine which combinations result in DCBH formation in the target atomic cooling halo and which do not. As an example, we show in Figure~\ref{fig.ratecurve} the full range of values of k$_{\rm de}$ and k$_{\rm di}$ we obtain with the SB99 SED grid (gray shaded region) and the BPASS SED grid (blue shaded region) for a halo-galaxy separation of 5~kpc. We see that for many combinations of stellar mass and stellar age, the LW flux reaching the atomic cooling halo is insufficient to enable DCBH formation, but that there are combinations of parameters that do yield a sufficiently large k$_{\rm di}$ (Eq.~\ref{eq.ratecurve}).}

%Additionally, \cite{InayoshiTanaka2014} have shown the dependence of $\Jc$ on the presence of soft X-rays. A large fraction of binary stellar populations, such as the one considered in this study, can produce a considerable X-ray emission as well. Thus, we include a discussion of our results in the light of these studies.
{The inclusion of binaries could have important consequences on the LW escape fraction as a recent study by \citet{2017arXiv170107031S} demonstrates its sensitivity to SED dependent quantities such as ionising radiation, and subsequently, self--shielding of H$_2$ by H. The assumption of a non-zero binary fraction also has an important consequence for the X-ray
 SED of the considered stellar populations. If the binaries in these systems have a similar
 distribution of separations to those in local star-forming galaxies, then at least some will
 eventually become high-mass X-ray binaries. High-mass X-ray binaries are the primary source 
 of X-rays in star-forming systems that lack an active galactic nucleus, and there is a
 well-established correlation between the star formation rate (SFR) and the X-ray luminosity ($L_{\rm X}$)
 of these systems \citep[see e.g.][]{gb03,mineo14}. If the high-redshift galaxies responsible for producing 
 the LW photons share this same correlation, then this can have important implications for the
 resulting value of $J_{\rm crit}$ \citep{io11,it15,Latif15,regan16,gl16}. We note however
 that the question of whether or not high redshift galaxies do show the same correlation between
 SFR and $L_{\rm X}$ remains unanswered, and the size of their impact on $J_{\rm crit}$ remains
 uncertain. In view of this, we do not account for the presence of X-rays in our current calculations. 
 In Sec. 3 we briefly discuss the possible impact of X-rays on our results.}
%%%%%%%%%%%%%%%%%%%%%%%%%%%%%%%%%%%%%%%%%%
\section{Results}
%%%%%%%%%%%%%%%%%%%%%%%%%%%%%%%%%%%%%%%%%%

We first plot the LW {output} and rate parameters from BPASSv2 and SB99 models in the top panel of Fig. 2. As expected, the LW output of BPASSv2 is higher than that of SB99 at ages $> 10 \rm \ Myr$. Considering this fact alone, one would expect the $\Jc$ from binary populations to be lower than the one from single stellar populations. However, the rate parameters, $\beta$ (middle panel) and $\alpha$ (bottom panel), for BPASSv2 are consistently lower than the ones produced by SB99 at all ages. This hints towards a more complicated interplay of the rates and the LW output leading to the need for a more in depth analysis of $\Jc$.  {In order to facilitate comparison with previous studies, we note here that a BPASSv2 galaxy with $\rm M_{*}=10^6 \msun$ and age$= 10$Myr has $\alpha / \beta \sim 0.5$ (See Fig. 2), which corresponds to a black body temperature of $3\times10^4\rm \ K$ \citep{Sugimura:2014p3946}.}

{In Fig.~\ref{fig.bpass}, we compare the results of our analysis from the SB99 SEDs (left) vs. BPASSv2 SEDs (right). We show the region in the M$_{\star}$--age parameter space in which DCBH formation is permitted (grey), where the labelled contours indicate various different values of $\Jlw$. The figure is split in top, middle and bottom panels corresponding to separations of 5, 12 and 20 kpc.} %In the right panels, we show the distribution of $\Jc$ obtained by lowering the $\Jlw$ in the grey regions of the left panel, till a minimum value of k$_{\rm di}$ that satisfies Eq.~\ref{eq.ratecurve} is obtained. The histograms are split by a stellar age of 400 Myr which roughly corresponds to the age of the Universe at $z=12$, and the  solid, dotted and dashed lines correspond to a separation of 5, 12 and 20 kpc respectively.
We find that the BPASS models produce systematically higher values of $\Jlw$ for any given combination of M$_*$ and age, particularly when $\rm M_{*}$ and the age are both large. For example, a galaxy with an age, t$_{*} = 10^{7.5}$~yr, a stellar, mass M$_{\star}\sim 10^{9.5}\msun$ and a separation of 5~kpc from the atomic cooling halo of interest produces $\Jlw \sim 700$ with the BPASSv2 model,  but only $\Jlw \sim 100$ with the SB99 model. This is because binary stellar populations yield more LW flux per stellar baryon especially at ages $\gtsim 10$ Myr (Fig.~\ref{fig.ratescompare}). {Therefore, particularly at late times, one would expect them to be more effective in producing a higher $\Jlw$ value at a given distance than single stellar populations.  {From this one would naturally infer that DCBH can occur more easily in the vicinity of binary populations, than in the vicinity of single stellar populations. However, we find that the $\Jc$ that is required for DCBH formation is higher from binaries than when we assume that all stars are single.}This  result is actually just a reflection of the fact that the value of $\Jc$ required for DCBH formation depends on the whole of the SED. Although BPASSv2 has a higher LW output, SB99 SEDs produce more lower energy photons and are thus much more effective at destroying H$^{-}$, as can be seen in Fig.~\ref{fig.ratescompare} where the values of $\alpha$ and $\beta$ plateau for BPASSv2 but steadily rise for SB99 at stellar ages $\gtsim$ 10 Myr. Consequently, with the SB99 SEDs, we require fewer LW photons in order to successfully suppress H$_{2}$ formation, and hence obtain a smaller $J_{\rm crit}$. 

%%%%%%%%%%%%%%%%%%%%%%%%%%%%%%%%%%%%%%%%%%
\begin{figure}
\includegraphics[width=0.75\columnwidth,angle=90,trim={-1cm 0cm 0cm -1.5cm},clip]{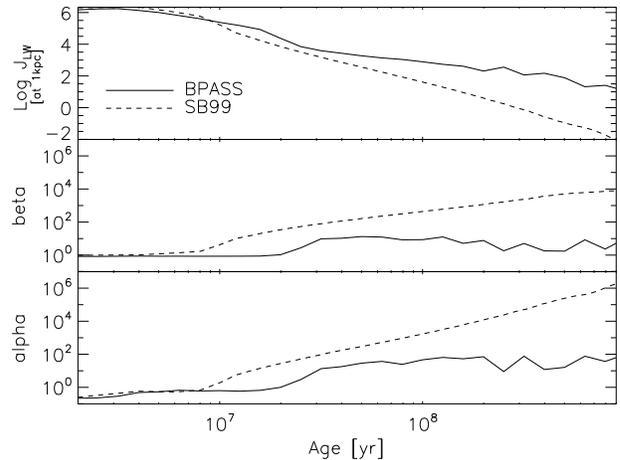}
\caption{The J$_{\rm LW}$ computed at 1 kpc for $\rm M_{*}= 10^6 \msun$ (top), and the rate parameter for 
H$_2$ photodissociation $\beta$ (middle) and H$^-$ photodetachment $\alpha$ (bottom) as a function of time using BPASSv2 and SB99.}
\label{fig.ratescompare}
\end{figure}
%%%%%%%%%%%%%%%%%%%%%%%%%%%%%%%%%%%%%%%%%%
%%%%%%%%%%%%%%%%%%%%%%%%%%%%%%%%%%%%%%%%%%
\begin{figure*}
\includegraphics[width=0.8\columnwidth,angle=90,trim={0.25cm 1cm 1cm 1cm},clip]{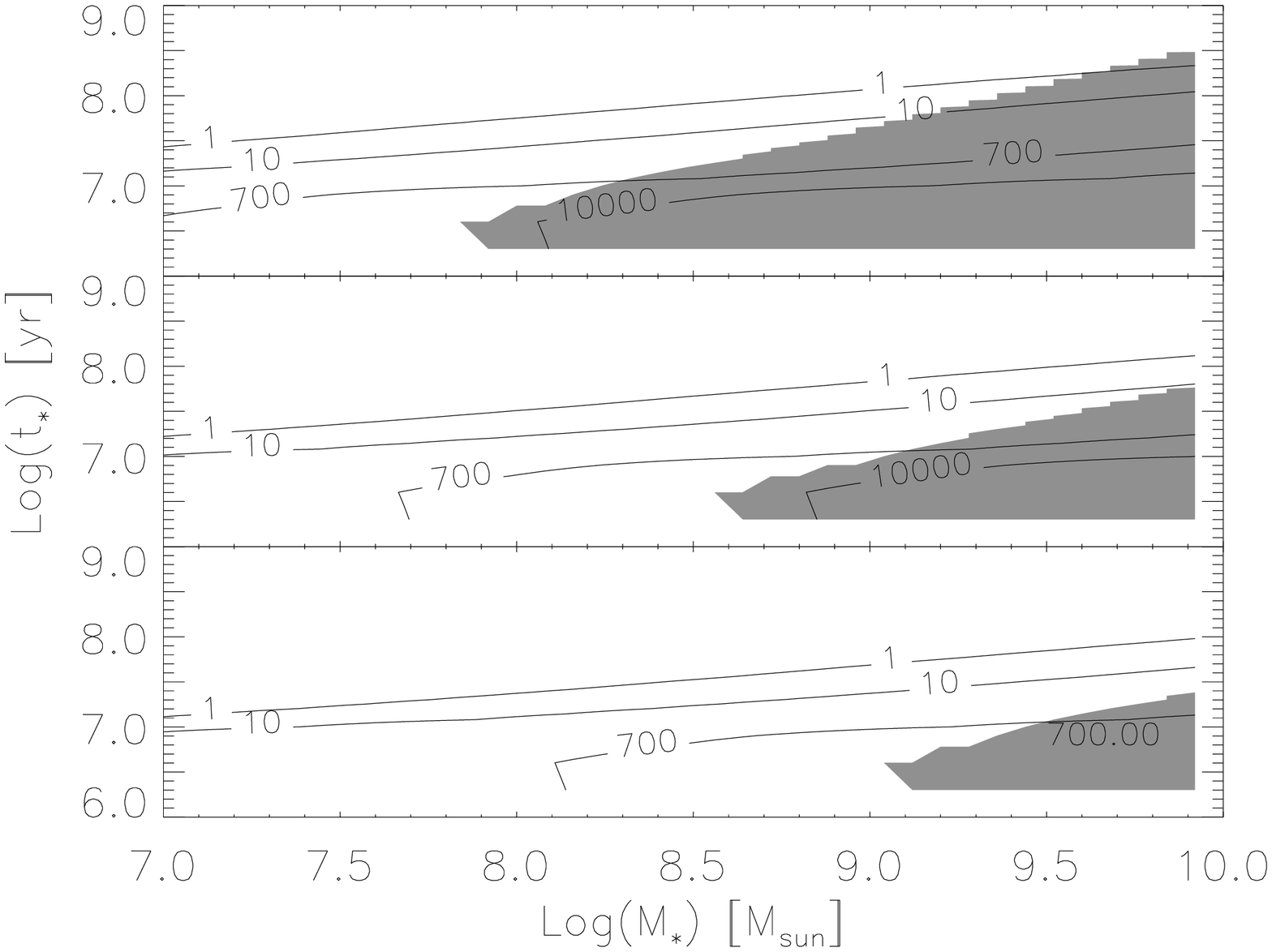}%\includegraphics[width=0.8\columnwidth,angle=90,trim={-1cm 0cm 0cm -0.2cm},clip]{ba_P5_grid_smass_Jcrit_st99_samebin_bpass.eps}\\
\includegraphics[width=0.8\columnwidth,angle=90,trim={0.25cm 1cm 1cm 1cm},clip]{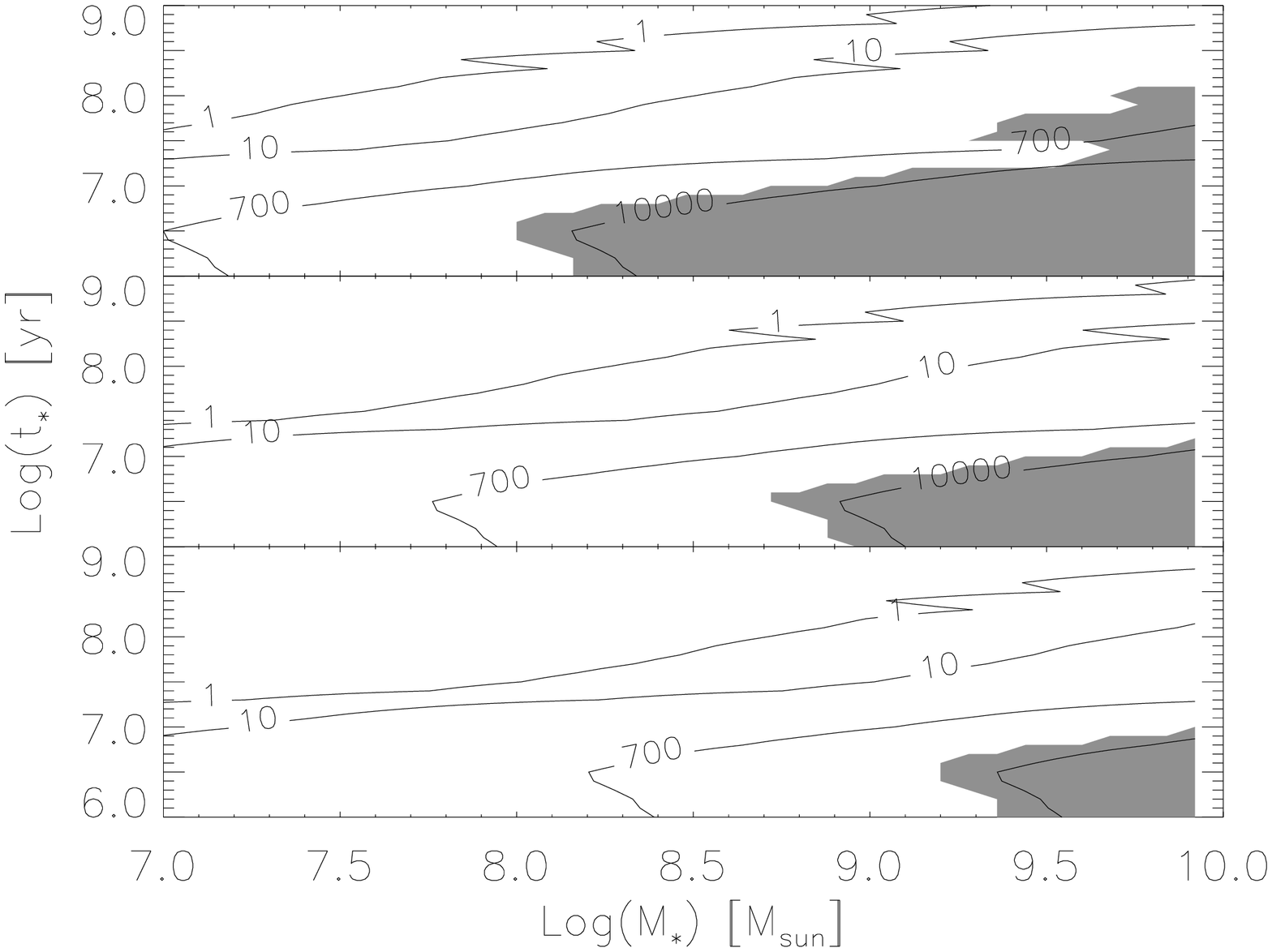}%\includegraphics[width=0.8\columnwidth,angle=90,trim={-1cm 0cm 0cm -0.2cm},clip]{ba_P5_grid_smass_Jcrit_bpass.eps}
\caption{Stellar populations that allow for DCBH formation, from SB99 shown on the right (taken from the Appendix of A16), and BPASSv2 on the right. Grey regions bound the $\rm M_*-\rm Age$ parameter space for which the stellar populations produce an H$_2$ photodissociation rate that at the location of the atomic cooling halo that satisfies Eq.~\ref{eq.ratecurve}. The top, middle and bottom panels are computed for an assumed separation of 5, 12 and 20 kpc between the atomic cooling halo and the irradiating source. The contours of $\Jlw$ at the respective distances are over-plotted in each of the panels.} %\textit{Right}: Histograms of $\Jc$ obtained by lowering the value of $\Jlw$ till Eq.~\ref{eq.ratecurve} is still valid in the grey regions from the left panel. The solid, dotted and dashed lines correspond to the 5, 12 and 20 kpc separations respectively and the panels are split by stellar age of 400 Myr which roughly corresponds to the age of the Universe at $z=12$.}
\label{fig.bpass}
\end{figure*}
%%%%%%%%%%%%%%%%%%%%%%%%%%%%%%%%%%%%%%%%%%

%, as can be understood by the values of $\alpha\ \& \ \beta$ that the BPASSv2 and SB99 SEDs produce. Although BPASSv2 has a higher LW output, SB99 SEDs produce far larger values of the rate parameters that govern Eqs.~\ref{eq.destroy}. This result is also seen in the form of the grey regions spanning a smaller range in the left panel of Fig.~\ref{fig.bpass} than in the left panel of Fig. B1 in A16. 
%Wolcott--Green et al. 2012 already hinted at this in their study when they found that the inclusion of 0.76 eV photons, or an Infra--Red background in the early universe can induce negative feedback on first star formation. 
%While both A16 and Sugimura et al. 2015 showed the importance of assumed SED slopes on J$_{crit}$, we further emphasize here the dependance of J$_{crit}$ on stellar populations older than 10 Myr.

{Further confirmation of this finding comes if we compare the distribution of $\Jc$ (for all three separations) in the Fig.~\ref{fig.jcrit_compare}. For the BPASSv2 SEDs, we find values in the range $\rm 100\ltsim J_{crit} \ltsim 3000$, depending on the age of the stellar population, whereas for SB99, the same IMF yields a much wider distribution with $\rm 0.1\ltsim J_{crit} \ltsim 3000$. 
%This is the critical effect of lower values of $\alpha\ \&\ \beta$ produced by older BPASS models than SB models. %In other words, at later times binary stellar populations are unable to produce enough H$^-$ photodetaching photons, thus leading to a drastic fall in k$_{de}$ which can not be countered by the higher LW flux. 
Although the curves are similar at ages $<$ 10 Myr, at later times, the $\Jc$ from binary populations is higher than the one required form single stellar populations. For example, at an age of 50 Myr, $\Jc \sim 100$ for the BPASSv2 SEDs, while it is only $\sim 10$ when derived using the SB99 SEDs. In fact, we see from Fig. 3 that a galaxy with $\rm M_* \gtsim 10^9$ and same age (50 Myrs) can easily have $\Jlw> \Jc$ when it is described by a SB99 SED, while for BPASSv2 SEDs $\Jlw<\Jc$ at this age for all masses.}

% {In Fig.~\ref{fig.jcrit_compare}, at an age of 50 Myr, $\Jc$ from BPASSv2 is $\sim 100$, whereas from SB99 models is $\sim 10$. From the left panels of Fig.~\ref{fig.bpass}, we see that at a distance of 5 kpc, a $\rm M_* \gtsim 10^9 \msun$ galaxy with the same age described by a SB99 SED can easily produce a $\Jlw>\Jc$, whereas for BPASSv2 $\Jlw<\Jc$ at all masses at $\rm t_*=50$ Myr.}

%This is further confirmed as the distribution of $\Jc$ in the right panel of Fig.~\ref{fig.bpass} is restricted to $\rm 100\ltsim J_{crit} \ltsim 3000$, whereas in A16, the same IMF yields a much wider distribution with $\rm 0.1\ltsim J_{crit} \ltsim 3000$. This is the critical effect of lower values of $\alpha\ \&\ \beta$ produced by older populations in BPASSv2 models than in SB99. %In other words, at later times binary stellar populations are unable to produce enough H$^-$ photodetaching photons, thus leading to a drastic fall in k$_{de}$ which can not be countered by the higher LW flux. 
%For all three separations, we plot the $\Jc$ from BP and SB in Fig.~\ref{fig.jcrit_compare}. The curves are similar at ages $<$ 10 Myr, after which the the $\Jc$ required from BPASSv2 is higher than the one required from SB, further supporting our hypothesis.\\

These findings lead us to conclude 

\begin{enumerate}
\item {\rm J$_{crit}$} does not solely depend on the LW photon yield, but on the 0.76 eV photon yield as well
\item The {\textit{distribution}} of {$\Jc$} depends on whether binaries are included in a galaxy's SED. 
For a stellar population of a given age and mass, the $\Jc$ is higher when binaries are considered.
\item The {\textit{distribution}} of {$\Jc$} is critically altered by the inclusion of older stellar populations. Our analysis shows that $\Jc$ originating from older single stellar populations ($> 10$ Myr) is much lower than the one from similarly aged binary stellar populations
\item Formation of DCBHs must be understood in terms of a critical region in the k$_{\rm de}$--k$_{\rm di}$ parameter space (Eq.~\ref{eq.ratecurve})
\end{enumerate}

{We note that point (i) is not a new result: it was already remarked upon by \citet{Sugimura:2014p3946} and in A16. However, our results here do help to} emphasize the dependance of J$_{\rm crit}$ on the shape of the SED, which {in turn} depends on physical parameters such as the inclusion of binaries and older stellar populations. 

%\begin{figure*}
%\includegraphics[width=0.85\columnwidth,angle=90,trim={0cm 0cm 0cm 0cm},clip]{ba_P5_grid_smass_contour_secondmodel.eps}\includegraphics[width=0.8\columnwidth,angle=90,trim={-1cm 0cm 0cm 0cm},clip]{ba_P5_grid_smass_Jcrit_st99_samebin_bpass.eps}
%\caption{SB99: contour and Jcrit}
%\label{fig.sb99}
%\end{figure*}

%%%%%%%%%%%%%%%%%%%%%%%%%%%%%%%%%%%%%%%%%%
\begin{figure}
\centering
\includegraphics[width=0.7\columnwidth,angle=90,trim={0cm 0cm 1cm 1cm},clip]{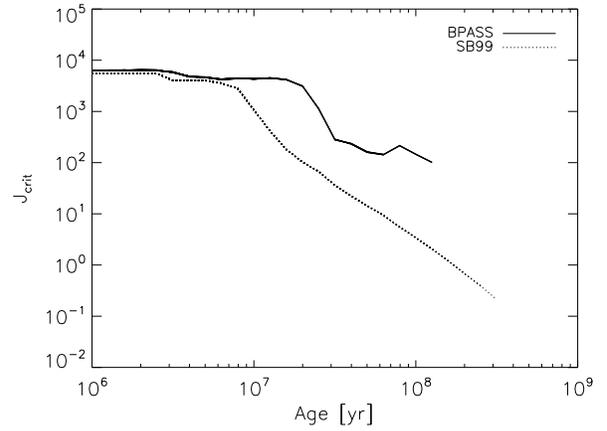}
\caption{Comparison of $\Jc$ for BPASSv2 (solid) and SB99 (dotted), for all separations. This is an age distribution of the histograms, for the entire separation range, shown in the right panel of Fig.~\ref{fig.bpass}.}
\label{fig.jcrit_compare}
\end{figure}
%%%%%%%%%%%%%%%%%%%%%%%%%%%%%%%%%%%%%%%%%%
High-redshift star-forming galaxies may also be bright at
 X-ray wavelengths leading to X-ray photoionization of the gas that produces additional free electrons.
This could lead to enhanced H$_{2}$ formation and hence X-rays can partially
 counteract the negative feedback due to LW photons and the softer optical and near-IR photons
 that destroy H$^{-}$ \citep[see e.g.][]{hrl96,gb03}. The impact of X-rays on $J_{\rm crit}$ has 
 been studied by several different authors \citep{io11,it15,Latif15,gl16}, but disagreements
 remain in the size of the overall effect. \citet{it15} find that if the incident LW spectrum is
 approximately described by a $T = 3 \times 10^{4}$~K black-body spectrum, then the value of
 $J_{\rm crit}$ in the presence of X-rays is given by
 \begin{equation}
 J_{\rm crit} = J_{\rm crit, 0} \left(1 + \frac{J_{\rm X, 21}}{2.2 \times 10^{-3}} \right)^{0.56},
 \end{equation}
 where $J_{\rm crit, 0}$ is the value of $J_{\rm crit}$ in the absence of X-rays and $J_{\rm X, 21}$
 is the strength of the X-ray background in units of $10^{-21} \: {\rm erg \, s^{-1} \, cm^{-2} \,
 Hz^{-1} \, sr^{-1}}$, measured at an energy of 1~keV. They also argue that $J_{\rm X, 21} \simeq
 4.4 \times 10^{-6} J_{\rm LW}$. Looking at the distribution of
 $J_{\rm crit}$ in our models shown in Fig. 4, we see that the largest values obtained are around $J_{\rm crit}
 \simeq 4000$. If this value is actually boosted by X-rays according to the \citet{it15}
 prescription, this would change the value to $J_{\rm crit, X} \simeq 14000$. 
 However, we note that other recent studies report a smaller effect. For example, \citet{gl16} finds an enhancement in $J_{\rm crit}$ that is
 roughly a factor of two smaller than in \citet{it15} at any given $J_{\rm X, 21}$, due to differences in the assumption 
 made regarding the effectiveness of X-ray shielding
 in the target halo. In that case, accounting for X-rays would increase our largest values of 
 $J_{\rm crit}$ by less than a factor of two, and hence would not significantly change the 
 qualitative results of our study. \citet{Latif15} find an even smaller effect, even with very
 large values of $J_{\rm X, 21}$ barely affecting the values of $J_{\rm crit}$ in their three--dimensional runs. In view of this uncertainty in the overall impact of X-rays, we neglect
 them in our current study, although we hope to return to this point in future work.

 {Recently \citet{Chon16a} studied the effects of tidal disruption of the DCBH host by the neighbouring galaxy responsible for the LW radiation field. They found that unless the DCBH host halo assembles via major mergers, it is prone to tidal disruption by the neighbouring galaxy. Thus if one interprets a higher J$_{crit}$ from binaries as a indication of a high stellar mass, then it is likely that tidal disruption events could render the neighbouring atomic cooling halo unsuitable for DCBH formation.}

%%%%%%%%%%%%%%%%%%%%%%%%%%%%%%%%%%%%%%%%%%
\section{Summary}
%%%%%%%%%%%%%%%%%%%%%%%%%%%%%%%%%%%%%%%%%%

We study the LW flux requirement for DCBH formation from galaxies that have a stellar population that includes a significant binary fraction. We show that despite their high LW output, binary populations are in fact inefficient at causing DCBH in their vicinity when compared to single stellar populations, contrary to what one would naively expect. This can be attributed to the SEDs of binary populations that are systematically bluer than those of populations composed only of single stars, meaning that the light from them is much less effective at causing H$^{-}$ photodetachment. The lower H$^{-}$ photodetachment rates mean that higher H$_{2}$ photodissociation rates are needed in order to bring about DCBH formation, and so the required values of $\Jc$ are larger. 

% {X-rays from binary stellar populations, such as the ones considered in this work could further have an effect on $\Jc$. For binary stellar populations less then 10 Myr old, a $\Jc \sim 4000$ implies an X-ray flux that boosts the $\Jc$ requirement to $\sim 60 \times 10^4$ \citep{InayoshiTanaka2014}. Following their Fig. 5, this implies that the number of DCBH sites is far less then the number of observed SMBHs at $z>6$. Although their calculation is based on drawing from a probability distribution function of pristine atomic cooling haloes exposed to a LW flux is based on a two-point correlation term, it provides important insights into the physical parameter space most relevant for DCBH formation.}

Consistent with A16, we a find a distribution in the values of the $\Jc$ produced by binary populations, albeit narrower ($\Jc \sim 300-3000$) than the one produced by single stellar populations ($\Jc \sim 0.1- 3000$). Furthermore the need for older single stellar populations becomes clear as they produce the lowest values of $\Jc$ in both cases, due to a higher k$_{\rm de}$. 
%{In this work, we assume that stellar populations are represented by a single burst. However accounting for galaxies' entire star formation history (SFH) can lead to a complex scenario. For example in the case of a SFH characterised by a burst at early times followed by an exponential decline of the star formation activity, it unclear which SED model would favour DCBH formation in a neighbouring halo at later times (Agarwal et al. in prep).}
% {In this work, galaxies were represented by a burst mode of star formation. However accounting for galaxies' entire star formation history (SFH) can lead to a complex scenario. For example consider a SFH characterised by a burst at early times followed by an exponential decline of the star formation activity. In this case, if the H$^-$ photodetachment rate is not high enough at a later time, DCBH formation in a neighbouring pristine atomic cooling halo might occur only when binaries are accounted for in the SED of the irradiating galaxy. This is because the presence of binaries would boost the overall UV emission at later times, leading to a high rate of H$_ 2$ photodissociation that can compensate for the low value of k$_{\rm de}$. In case of SB99, the dearth of UV photons could cause $\Jlw < \Jc$.}
This pushes the idea further that the formation of DCBHs must be understood in terms of the k$_{\rm de}$--k$_{\rm di}$ parameter space (Eq.~\ref{eq.ratecurve}), and not in terms of a single flux value. 
% {As one would expect, lowering the binary fraction would push the BPASSv2 curve in Fig.~\ref{fig.jcrit_compare} closer to SB99. However, the effect of a higher binary fraction is unclear as the excess of UV photons could lead to a high enough photodissociation rate of H$_2$ (i.e. higher k$_{\rm di}$), which could compensate for the low of H$^-$ photdetachment rate (i.e. low k$_{\rm de}$).}

%%%%%%%%%%%%%%%%%%%%%%%%%%%%%%%%%%%%%%%%%%%%%%%%%%%%%%%%%%%%%%%%%%%%%%%%%%
%%%%%%%%%%%%%%%%%%%%%%%%%%%%%%%%%%%%%%%%%%%%%%%%%%%%%%%%%%%%%%%%%%%%%%%%%%
\section*{Acknowledgements}

BA would like to thank Laura Morselli, Eric Pellegrini and Claes-Erik Rydberg for useful discussions. BA, RSK, SCGO would like to acknowledge the funding from the European Research Council under the European Community's Seventh Framework Programme (FP7/2007-2013) via the ERC Advanced Grant STARLIGHT (project number 339177).
Financial support for this work was provided by the Deutsche Forschungsgemeinschaft via SFB 881, "The Milky Way System" (sub-projects B1, B2 and B8) and SPP 1573, "Physics of the Interstellar Medium" (grant number GL 668/2-1).
%\adr{Fill in here}

\bibliographystyle{mn2e}
\bibliography{babib}
\label{lastpage}
\end{document}